\documentstyle [12pt,eqsecnum,aps,amsfonts,axodraw] {revtex}
\tighten \draft
\input epsf
\newcommand{\beq}{\begin{equation}}
\newcommand{\eeq}{\end{equation}}
\newcommand{\beqs}{\begin{eqnarray}}
\newcommand{\eeqs}{\end{eqnarray}}

\begin{document}

\draft

\begin{flushright}
{UT-03-31, \ YITP-03-50}
\end{flushright}

\baselineskip 6.0mm

\vspace{6mm}

\begin{center}
{\Large{\bf On a Neutrino Electroweak Radius}}
\end{center}
\vskip .5 truecm
\centerline{\bf Kazuo Fujikawa }
\vskip .4 truecm
\centerline {\it Department of Physics,University of Tokyo}
\centerline {\it Bunkyo-ku,Tokyo 113,Japan}
\vskip 0.5 truecm
\centerline{\bf Robert Shrock} 
\vskip .4 truecm 
\centerline{\it C. N. Yang Institute for Theoretical Physics} 
\centerline{\it State University of New York, Stony Brook, NY 
11794 USA} 
\vskip 0.5 truecm 

\begin{abstract}

We study a combination of amplitudes for neutrino scattering that can isolate a
(gauge-invariant) difference of chirality-preserving neutrino electroweak radii
for $\nu_\mu$ and $\nu_\tau$.  This involves both photon and $Z_\mu$
exchange contributions.  It is shown that the construction singles out the
contributions of the hypercharge gauge field $B_{\mu}$ in the standard model.
We comment on how gauge-dependent terms from the charge radii cancel with
other terms in the relative electroweak radii.

\end{abstract}


\vspace{16mm}

\pagestyle{empty}
\newpage

\pagestyle{plain} \pagenumbering{arabic}
\renewcommand{\thefootnote}{\arabic{footnote}}
\setcounter{footnote}{0}

\section{Introduction}

Electromagnetic properties of neutrinos are of fundamental importance and serve
as a probe of whether neutrinos are Dirac or Majorana particles and of new
physics beyond the standard model (SM) \cite{bardeen}-\cite{ncr}.  In
general, the matrix element of the electromagnetic current between initial
and final neutrinos $\psi_j$ and $\psi_k$ with 4-momenta $p$ and $p^\prime$ is
\beqs
\langle \psi_k(p^\prime) | J_\lambda | \psi_j(p) \rangle & = &
e \ \bar \psi_k(p^\prime)\biggl \{ \gamma_\lambda 
\Bigl [ F_1^V(q^2)+F_1^A(q^2)\gamma_5 \Bigr ] 
+i \frac{\sigma_{\lambda \rho}q^\rho}{m_{\nu_k}+m_{\nu_j}} \Bigl [ F_2^V(q^2)
+ F_2^A(q^2)\gamma_5 \Bigr ] \cr\cr
& & + q_\lambda \Bigl [ F_3^V(q^2) + F_3^A(q^2)\gamma_5 \Bigr ] 
\biggr \} \psi_j(p) \ .
\label{matrixelement}
\eeqs
where $e$ is the electromagnetic coupling and $q=p-p^\prime$.  The form factors
$F_n^{V,A}$ are matrices in the space of neutrino mass eigenstates, and their
$(kj)$ elements appear in the above amplitude \cite{cc}.  In the diagonal case
$j=k$, \ $e F_2^V(0)_{jj}/(2m_{\nu_j})$ is the magnetic dipole moment of the
mass eigenstate $\nu_j$ and $-ie F_2^A(0)_{jj}/(2m_{\nu_j})$ gives the electric
dipole moment.  A Dirac neutrino in the standard model generalized to include
such masses has a magnetic moment $\mu_{\nu_j} = 3eG_F m_{\nu_j}/(8 \pi^2
\sqrt{2})$ \cite{fs}, while in models with right-handed charged currents, this
quantity also involves terms depending on charged lepton masses
\cite{jkim,bmr}.  A Dirac neutrino may also have a CP-violating electric dipole
moment (e.g. \cite{rs82}).  For a Majorana neutrino, with $\psi_j = \psi_j^c$,
these operators vanish identically, so that $\mu_{\nu_j} = d_{\nu_j}=0$.

It is also of interest to consider the chirality-preserving terms $\bar\psi_k
\gamma_\lambda \psi_j$ and $\bar\psi_k \gamma_\lambda \gamma_5 \psi_j$ in
eq. (\ref{matrixelement}) and the associated form factors $F_1^V(q^2)$ and
$F_1^A(q^2)$.  Although the electric neutrality of the neutrino means that
$F_1^V(0)_{jj}=0$, one may consider the Taylor series expansions of
$F^{V,A}_1(q^2)_{jj}$ as functions of $q^2$, in particular,
\beq
\langle r^2 \rangle_{\nu_j} =
6 \frac{d F_1^V(q^2)_{jj}}{dq^2}|_{q^2=0} \ . 
\label{rsqgen}
\eeq
This is often called the neutrino charge radius, and we shall follow this
convention; more precisely, it is the charge radius squared.  By itself, the
neutrino charge radius is gauge-dependent and hence is not a physical
observable \cite{bardeen,ls} (contrary to the recent claim in
Refs. \cite{bern2,bern3}).  Explicit calculations in unified renormalizable
electroweak gauge theories \cite{ls} using the $R_\xi$ gauge
\cite{fujikawa,fls} clearly displayed the gauge dependence of various
quantities including the charge radius.

It is useful to consider the construction of a gauge-independent set of
amplitudes involving the chirality-preserving terms in
eq. (\ref{matrixelement}) that can serve to characterize neutrino properties.
In this paper we shall discuss the construction of a set obtained from
differences of neutrino scattering amplitudes, giving details of our note
\cite{ncr} and how this relates to, and differs from, the recent approach of
Refs. \cite{bern1}-\cite{bern3}.  Since our focus here is on constraints from
gauge invariance and since for the chirality-preserving terms under
consideration, neutrino masses do not play as important a role as they do in
the chirality-flipping terms in (\ref{matrixelement}), we shall make the
simplification of working within the standard model with massless neutrinos.
(These simplifications were also made in Refs. \cite{bern1}-\cite{bern3} and
\cite{ncr}.)  For this massless neutrino case, Dirac and Majorana neutrinos are
equivalent, and there is no lepton mixing, so that the neutrino mass and group
eigenstates coincide.  From eq. (\ref{matrixelement}) the relevant matrix
element is then
\beq
\langle \psi_j(p^\prime) | J_\lambda | \psi_j(p) \rangle = 
e \ \bar \psi_j(p^\prime) \gamma_\lambda (1-\gamma_5) F_1(q^2)_j \psi_j(p) \ , 
\label{matrixelement_simp}
\eeq
where $F_1^V(q^2)_{jj} = F_1^A(q^2)_{jj} \equiv F_1(q^2)_j$.  (We will often
drop the subscript $j$ where it is obvious from the context.)

\section{Neutrino Scattering Reactions} 

\begin{center}
\begin{figure}[]
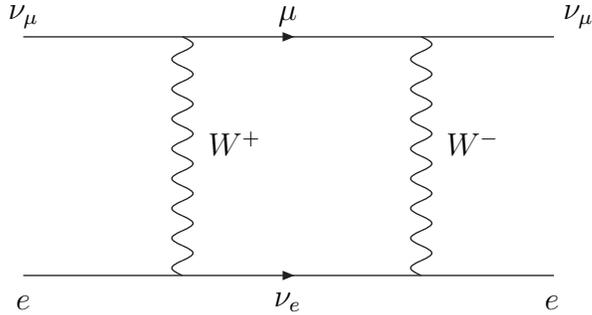

\caption{\footnotesize{$2W$-exchange graph, whose contribution
vanishes for $e=e_R$ in the SM.}}
\label{2W-graph}
\end{figure}
\end{center}

To begin, consider the tree and one-loop scattering amplitudes for the reaction
\beq
\nu_{\mu}(p) + e(k) \to \nu_{\mu}(p^{\prime}) + e(k^{\prime}) 
\label{reaction}
\eeq
(where all the particles appearing in the above process are physical, on-shell
particles).  The four-momentum transfer squared is denoted $t=q^2$ and the
center-of-mass energy squared is denoted $s$. Consider next the high-energy
limit $s/m_e^2 >> 1$ and specialize to the amplitude for the reaction 
\beq
\nu_{\mu}(p)+ e_{R}(k) \to \nu_{\mu}(p^{\prime})+ e_{R}(k^{\prime}) \ . 
\label{reaction_er}
\eeq 
where $e_{R}=P_Re$ denotes a right-handed electron (to be precise the helicity
plus electron which does not coincide with the right-handed electron specified
by $P_{R}$ in the presence of the non-vanishing electron mass), where $P_{R,L}
= (1/2)(1 \pm\gamma_5)$ are chirality projection operators.  The tree-level
diagram for this reaction involves the exchange of a $Z$ boson in the
$t$-channel.  An important simplifying approximation is that the electron mass
is neglected for this high-energy limit even for $q^2=0$, except for an
infinitesimal electron mass on internal fermion lines to control infrared
divergences.  With this approximation of neglecting the electron mass, so that
a positive-helicity electron is equivalent to a right-handed electron, the
contribution of the box diagram with $2W$ exchange, shown in
Fig. \ref{2W-graph}, vanishes.  As a result, one can extract the contributions
from the one-photon and one-$Z$ exchange diagrams with one-loop self-energy and
vertex corrections together with $2Z$ exchange box diagrams, shown in the
appendix.  This amplitude with right-handed massless electrons is
gauge-independent and constitutes a part of the physical S-matrix.

\bigskip
\bigskip
\bigskip

\begin{center}
\begin{picture}(210,100)(0,0)
\ArrowLine(0,100)(200,100)
\ArrowLine(0,10)(200,10)
\Photon(60,10)(60,100){4}{8}
\Photon(150,10)(150,100){4}{8}
\Text(0,108)[]{$\nu_\mu$}
\Text(100,108)[]{$\mu$}
\Text(210,108)[]{$\nu_\mu$}
\Text(80,60)[]{$W^+$}
\Text(170,60)[]{$W^-$}
\Text(0,0)[]{$e$}
\Text(100,0)[]{$\nu_e$}
\Text(200,0)[]{$e$}
\end{picture}
\end{center}

We remark on a related but different approach to the analysis of the matrix
element. In Refs. \cite{bern1}-\cite{bern3}, the authors write the Feynman
diagrams defined in the $R_{\xi}$ gauge and then apply the ``pinch technique'',
previously discussed in \cite{cornwall,degrassi}, to the complete set of
Feynman diagrams contributing to this S-matrix element.  This technique
involves a rearrangement of various Feynman diagrams for an element of the
physical S-matrix before the calculation of loop integrals. For a tree-level
amplitude contributing to a physical S-matrix element, it is trivial to
redefine the Feynman rules given by the $R_{\xi}$ gauge to those of the
background Feynman gauge.  If one uses dimensional regularization, for example,
all the Feynman diagrams are finite at the one-loop level also, and thus the
rearrangement of Feynman diagrams is justified.  The starting Feynman rules are
identical in the conventional formulation and with this pinch technique, and
thus the result is also identical, provided that the method is well-defined.
   
After this rearrangement, one obtains the amplitudes written in terms of the
Feynman gauge in the background field method\cite{dewitt}-\cite{grassi},
\cite{monyonko}, which exhibits the $U(1)_{em}$ symmetry of the electromagnetic
interaction explicitly.  This rearrangement of various Feynman diagrams and the
cancellation of the gauge parameter $\xi$ for the physical process is
consistent with the general formulation of gauge theory. It therefore follows
that physical results obtained via the conventional formulation of gauge
theories and via the use of this pinch technique must be the same, since the
Feynman gauge in the background field method is one of the allowed gauge
conditions. However the Feynman gauge in the background field method has no
privileged position among various possible gauge conditions.  For example, the
$U(1)_{em}$ symmetry of the electromagnetic interaction remains intact in a
non-linear $R_{\xi}$ gauge\cite{fujikawa} also.

\section{Method Using Differences of Neutrino Reactions} 

One way to separate the contributions of the $2Z$ exchange box diagrams from
those of the one-photon and one-$Z$ exchanges involves the
``neutrino-antineutrino method'' utilized in Refs. \cite{bern2,bern3} (see
Ref. \cite{ssm} for earlier related work), in which one considers the sum
$d\sigma(\nu_\mu e_R \to \nu_\mu e_R)/dq^2 + d\sigma(\bar\nu_\mu e_R \to
\bar\nu_\mu e_R)/dq^2$, specifically the $q^2 \to 0$ limit \footnote{The
crucial Eq. (2) in Ref. \cite{bern2}, which is the basis of the analysis of
Refs. \cite{bern2,bern3}
\beq
\bar{v}(p_{1})\gamma_{\mu}P_{L}v(p_{2})
=-\bar{u}(p_{2})\gamma_{\mu}P_{R}u(p_{1}), 
\eeq
is however not justified, as can be confirmed by considering the time component
of the current for $p_1=p_2$; in the case, this relation leads to
\beq
||P_{L}v(p_{1})||^{2}=-||P_{R}u(p_{1})||^{2} \ 
\eeq
and thus $P_{L}v(p_{1})=P_{R}u(p_{1})=0$.}. 

The idea underlying the neutrino-antineutrino method is simplified and 
the basic physical idea is precisely stated in a gauge-independent way if
one compares the process (\ref{reaction_er}) to the process with the 
charge-conjugated electron (i.e., positron) (see also Ref. \cite{ssm})
\beq
\nu_{\mu}(p)+e^{+}_{L}(k)\rightarrow \nu_{\mu}(p^{\prime})+
e^{+}_{L}(k^{\prime}).
\label{conj_reaction}
\eeq
In the massless electron case, the only interaction of the right-handed
electron is given by
\beq
{\cal L}_{int}=-g^{\prime}\bar{e}_{R}\gamma^{\mu}B_{\mu}e_{R}
\label{Lint}
\eeq
where $B_{\mu}$ stands for the hypercharge gauge field 
associated with the $U(1)_{Y}$ factor of the standard model.  Recall that 
$B_{\mu}$ is a linear combination of the photon $A_{\mu}$ and $Z_{\mu}$ fields
satisfying 
\beq
g^{\prime}B_{\mu}=\frac{g^{\prime}}
{\sqrt{g^{2}+(g^{\prime})^{2}}}(gA_{\mu}
-g^{\prime}Z_{\mu})=eA_{\mu}-G\sin^{2}\theta_{W}Z_{\mu}
\label{brel}
\eeq
with $g$ and $g^{\prime}$, respectively, standing for the $SU(2)_{L}$ and
$U(1)_{Y}$ gauge coupling constants and $G=\sqrt{g^{2}+(g^{\prime})^{2}}$. If
one rewrites the interaction (\ref{Lint}) with the charge-conjugated variables
defined by
\begin{equation}
e(x)=-C^{-1}(\overline{e^c})^{T}(x), \ \ \ \bar{e}(x)=(e^{c})^{T}(x)C \ , 
\label{conjugaterel}
\end{equation}
it can be expressed equivalently (considering a suitable limit of the 
point-splitting definition of the current and the anti-commuting
property of the electron field) as 
\begin{equation}
{\cal L}_{int}=g^{\prime}\overline{e^c}_{L}\gamma^{\mu}B_{\mu}
e^{c}_{L}.
\label{otherint}
\end{equation} 
If one compares the above two processes (\ref{reaction_er}) and
 (\ref{conj_reaction}) with the identical kinematical configurations, one can
 distinguish the amplitudes with odd powers in $g^{\prime}B_{\mu}$ (in the
 electron sector) from the amplitudes with even powers in $g^{\prime}B_{\mu}$
 (i.e., the $2Z_{\mu}$ exchange box diagrams), except for the wavefunction
 $P_{L}u(k)$ for $e^{c}_{L}$ replacing $P_{R}u(k)$ for $e_{R}$, by noting that
\begin{equation} 
\langle e^{c}_{L}(x)\overline{e^c}_{L}(y)\rangle
=P_{L}\langle e^{c}(x)\overline{e^c}(y)\rangle P_{R}
=P_{L}\langle e(x)\bar{e}(y)\rangle P_{R}
\label{prel}
\end{equation}
in the Dyson expansion of the S-matrix. The appearance of the positive-energy
solutions in both cases,\footnote{We expand the generic Dirac field as
\begin{eqnarray}
\psi(x)=\sum_{s}\int\frac{d^{3}p}{\sqrt{(2\pi)^{3}2p_{0}}}
[u(p,s)b(p,s)e^{-ipx}+v(p,s)d^{\dagger}(p,s)e^{ipx}]
\nonumber
\end{eqnarray}
with the charge conjugation relations 
\begin{eqnarray}
C\gamma^{\mu}C^{-1}=-(\gamma^{\mu})^{T}, \ \ C\gamma_{5}C^{-1}
=\gamma^{T}_{5},\ \ v^{T}(p,s)C=\bar{u}(p,s), \ \ 
-C^{-1}\bar{v}^{T}=u(p,s),\ \ C^{\dagger}C=1,\ \ C^{T}=-C.
\nonumber
\end{eqnarray}
We then have 
\begin{eqnarray}
\psi^{c}(x)=-C^{-1}\bar{\psi}^{T}=
\sum_{s}\int\frac{d^{3}p}{\sqrt{(2\pi)^{3}2p_{0}}}
[v(p,s)b^{\dagger}(p,s)e^{ipx}+u(p,s)d(p,s)e^{-ipx}].
\nonumber
\end{eqnarray}
}
and the difference between $P_{L}u(k)$ and $P_{R}u(k)$ does not matter in the
evaluation of the forward cross section~\cite{bern2,bern3}. The
interference term of box diagrams with the $Z_{\mu}$-exchange tree diagram,
which is the relevant quantity in the lowest order process beyond the tree
process, thus changes sign between the above two processes. One can
thus eliminate the $Z_{\mu}$ box diagram contributions by a physical operation
by considering the sum of the two cross sections (in the limit $q^2 \to 0$) 
\begin{equation}
\frac{d\sigma(\nu_{\mu}+e_{R} \to \nu_{\mu}+e_{R})}{dq^2} + 
\frac{d\sigma(\nu_\mu+e^c_L   \to \nu_\mu+e^c_L)}{dq^2} \ . 
\end{equation}
That is, since $d\sigma/dq^2 = |A_{tree}|^2 + 2 Re(A_{tree}A_{1-loop}^*)$ plus
higher-order terms, and since for the the second term, $A_{tree}$ reverses sign
for the reaction with $e_R$ replaced by $e^c_L$ while the 2$Z$ exchange graphs
do not, this sum removes terms from the 2-$Z$ exchange box diagrams.  These
terms from the 2$Z$ box diagrams are gauge-independent by themselves, as can
be explicitly confirmed; the gauge parameter for $Z_{\mu}$ cancels among the
box and crossed diagrams (see also \cite{degrassi0,papavassiliou}). The
physical separation of the $2Z$ exchange contributions is thus perfectly
consistent with the basic principle of gauge theory.

This physical separation of these box diagrams is an important operation, but
it is clear that what one measures after this separation is the form factor of
the neutrino detected by the hypercharge gauge field $B_{\mu}$, which is a
linear combination of the physical fields $A_{\mu}$ and $Z_{\mu}$.  The
neutrino-anti-neutrino method with a massless electron gives a physical basis
for writing the gauge-independent one-loop amplitude (by excluding the $2Z$
box diagrams and neglecting the wave function renormalization factors for
simplicity)
\begin{eqnarray}
&&e^{2}\bar{u}_{R}\biggl [ \frac{F_1(q^{2})}{q^{2}}
+\frac{F^{\nu\nu}_{Z}(q^{2})
+F^{ee}_{Z}(q^{2})}{q^{2}-M^{2}_{Z}}
-\frac{\Pi_{ZZ}(q^{2})}{(q^{2}-M^{2}_{Z})^{2}}
-\frac{\Pi_{\gamma Z}(q^{2})}{q^{2}(q^{2}-M^{2}_{Z})} \biggr ]
\gamma^{\alpha}u_{R}\nonumber\\
&&\times  
\bar{\nu}_{\mu}\gamma_{\alpha}(1-\gamma_{5})\nu_{\mu}.
\label{amp}
\end{eqnarray}
Here $F_1(q^{2})$ and $F^{\nu\nu}_{Z}(q^{2})$ denote the $\nu_\mu$ vertex
functions measured by $B_{\mu}$, and $F^{ee}_{Z}(q^{2})$ is the electron vertex
function measured by $Z_{\mu}$. The self-energy corrections $\Pi_{ZZ}(q^{2})$
and $\Pi_{\gamma Z}(q^{2})$ stand for the two-point functions for
$B_{\mu}-Z_{\mu}$ coupling. The gauge independence of (3.10) is established by
a simple argument of the gauge independence of the physical S-matrix without
referring to the technical details such as the pinch technique or the
background field method.

\section{Relative Electroweak Radii}

One can further consider the difference of (\ref{amp}) and the corresponding 
amplitude with $\nu_{\mu}$ replaced by $\nu_\tau$
by measuring the difference~\cite{bern2,bern3}
\beqs
& & \biggl [ \frac{d\sigma(\nu_\mu+e_R \to \nu_\mu+e_R)}{dq^2} +
\frac{d\sigma(\nu_\mu+e^c_L \to \nu_\mu+e^c_L)}{dq^2} \biggr ] \cr\cr
& & - 
\biggl [ \frac{d\sigma(\nu_\tau+e_R \to \nu_\tau+e_R)}{dq^2} +
\frac{d\sigma(\nu_\tau+e^c_L \to \nu_\tau+e^c_L)}{dq^2} \biggr ] \ . 
\eeqs 
In this way one can eliminate the common terms $F^{ee}_{Z}(q^{2})$, \ 
$\Pi_{ZZ}(q^{2})$, and $\Pi_{\gamma Z}(q^{2})$ in the two amplitudes, since one
can measure the amplitude (\ref{amp}) and its $\nu_\tau$ analogue
themselves as interference terms with the leading tree-level $Z$ exchange
diagram.  One thus arrives at the physical quantity measured by the hypercharge
gauge field $B_{\mu}$
\begin{eqnarray}
e^{2}\lim_{q^2 \to 0} \biggl [ \frac{F_1(q^{2})}{q^{2}}
+\frac{F^{\nu\nu}_{Z}(q^{2})}{q^{2}-M^{2}_{Z}} \biggr ] |_{\nu_\mu}
\ - \ e^{2}\lim_{q^2 \to 0} \biggl [ \frac{F_1(q^{2})}{q^{2}}
+\frac{F^{\nu\nu}_{Z}(q^{2})}{q^{2}-M^{2}_{Z}}\biggr ] |_{\nu_\tau}
\label{fdif}
\end{eqnarray}
which is gauge-independent. The gauge independence of (\ref{fdif}) is
equivalent to the gauge independence of (\ref{amp}) and its $\nu_\tau$
analogue. However, the physical separation of the photon exchange
contributions from the $Z_{\mu}$ exchange contributions in (\ref{fdif}), namely
the separation of $F_1(q^{2})$ from $F^{\nu\nu}_{Z}(q^{2})$, is not obvious in
the standard model.  Indeed, previous detailed calculations~\cite{ls,lucio2} of
$\frac{F_1(q^{2})}{q^{2}}|_{q^{2}=0}$ show that the sub-leading term of the
order $m^{2}_{l}/M^{2}_{W}$ with $l=\mu \ {\rm or}\ \tau$ in the relative
charge radius (squared) 
\begin{equation}
\Delta \langle r^{2} \rangle =6 \lim_{q^2 \to 0}
\biggl [ \frac{F_1(q^{2})}{q^{2}}|_{\nu_\mu}-
\frac{F_1(q^2)}{q^{2}}|_{\nu_\tau} \biggr ] 
\label{rsqdif}
\end{equation}
depends on the gauge parameter in $R_{\xi}$ gauge and diverges as $\xi \to 0$
like 
\begin{equation}
 \frac{3g^{2}}{128\pi^{2}M^{2}_{W}}
\biggl [ \frac{m^{2}_{\mu}-m^{2}_{\tau}}{M^{2}_{W}} \biggr ]\ln \Bigl ( 
\frac{1}{\xi} \Bigr )
\label{gaugedep} 
\end{equation}
(see eqs.(2.30) and (2.54) in~\cite{ls} and also eq.(4a) in~\cite{lucio2},
which confirmed the calculation in \cite{ls}). This shows that a
gauge-independent separation of $F_1(q^{2})$ from $F^{\nu\nu}_{Z}(q^{2})$ in
(\ref{fdif}) is not possible, since if it were possible, the 
above gauge
parameter would not appear. The leading contribution to the relative charge
radius squared in (\ref{rsqdif}) (see, for example, \cite{ls,lucio2})
\begin{equation}
\Delta \langle r^{2} \rangle_{leading}=\frac{g^{2}}{16\pi^{2}M^{2}_{W}}
\biggl [ \ln\frac{M^{2}_{W}}{m^{2}_{\mu}}
-\ln\frac{M^{2}_{W}}{m^{2}_{\tau}} \biggr ]
\label{leading}
\end{equation}
is formally gauge-independent, but we emphasize that the separation of the
leading term from the sub-leading term is not well-defined for the
gauge-dependent relative charge radius, since the sub-leading term can be made
arbitrarily large by gauge choice, as is evident in eq. (\ref{gaugedep}).

The relative electroweak radius defined in (\ref{fdif}) as a combination of
$F_1(q^{2})$ and $F^{\nu\nu}_{Z}(q^{2})$ is gauge-independent, and only in this
combination can one separate the leading term from the sub-leading term.  Our
analysis of (\ref{leading}) is consistent with the result in
Ref.~\cite{bern2,bern3}, which obtains $\Delta F^{\nu\nu}_{Z}(0)=0$ up to terms
of the order \cite{renorm} $m^{2}_{l}/M^{2}_{W}$ with $l=\mu \ {\rm or}\ \tau$;
that is, the leading term of $\Delta F^{\nu\nu}_{Z}(0)=0$ vanishes and is thus
gauge independent.  The leading term of the gauge independent electroweak
radius in (\ref{fdif}) thus agrees with the value (\ref{leading}), which is, in
fact, gauge-independent.  The sub-leading term in $\Delta F^{\nu\nu}_{Z}(0)$ is
non-vanishing and is required to cancel the gauge dependence of the sub-leading
term in $\Delta F_1(q^{2})/q^{2}$ at $q^{2}=0$ in (\ref{fdif}), namely,
(\ref{gaugedep}).

We emphasize that the non-vanishing sub-leading term in $\Delta
F^{\nu\nu}_{Z}(0)$ is a manifestation of the gauge dependence of the relative
neutrino charge radius in the pinch technique; that is, one cannot eliminate
the $Z_{\mu}$ contamination in a gauge-invariant manner, since only the
combination (\ref{fdif}) is gauge-independent. The non-existence of a strictly
gauge-independent relative neutrino charge radius thus persists both in the
conventional formulation and with the pinch technique.  Our analysis in the
framework of conventional gauge theory thus clearly explains what is going on
in the complicated pinch technique analysis in~\cite{bern2,bern3}.  Moreover,
our analysis shows that the result for the gauge-independent relative neutrino
charge radius in~\cite{bern2,bern3} arises from the extra approximation of
neglecting the sub-leading terms and not from the use of the pinch
technique. If one neglects the sub-leading terms, one can readily establish the
gauge independence of the relative neutrino charge radius in the
conventional formulation also. The pinch technique does not produce any result
different from that of the conventional formulation.

Our analysis clearly shows that the relative neutrino charge radius is not
gauge-independent, much less the charge radius for an individual species of
neutrino.  As for the detailed analysis on the basis of the pinch technique
\cite{ncrit}, this may be useful to confirm the gauge independence of the
physical S-matrix element.  But one cannot infer that each part of the total
amplitude separately has a gauge-independent physical meaning simply
because the gauge parameter formally disappeared in the operation of the pinch
technique; the Feynman rules in the Feynman gauge would always be free of
gauge parameters in such a sense. Only the total amplitude for the S-matrix is
gauge-independent. Without the photon pole in the above neutrino scattering
process, no general principle can be used to argue for the gauge independence
of the one-photon exchange amplitude.  To establish the gauge independence of
the relative neutrino charge radius, one would need to restore the gauge
parameter in the photon exchange diagrams without changing the neutrino charge
radius, as is demonstrated for the case of the muon magnetic moment in
Ref.~\cite{fls}. This is, however, equivalent to the gauge parameter
independence of the vertex corrections to the photon exchange diagrams by
themselves before one applies the pinch technique.

Our analysis suggests that a useful quantity is the ``relative electroweak
radius'' in (\ref{fdif}) measured by the hypercharge gauge field
$B_{\mu}$ in the standard model.  This relative electroweak radius is
gauge-independent, and its leading term agrees with the leading term of the
relative neutrino charge radius in Refs.~\cite{bern1,bern2,bern3} defined by
the Feynman gauge in the background field method.  We believe that our
definition of the relative electroweak radius is conceptually simple and clear.

\section{Discussion and Conclusion}
 
The notion of the physical neutrino charge radius (squared) would be important
in the analysis of observables in gauge theory if this quantity were
gauge-invariant.  However, as was established long ago \cite{bardeen,ls}, it is
not gauge-invariant and hence is not a physical observable (contrary to the
recent claim in Refs. \cite{bern2,bern3}).  In this paper, in the limit
$s/m_e^2 >> 1$, we have constructed for the standard model a combination of
terms that provides a gauge-invariant quantity that may be regarded as a
relative electroweak chirality-preserving quantity, a sort of gauge-invariant
generalization of a neutrino charge radius.  We have shown that this involves
the hypercharge gauge field $B_{\mu}$ in a natural way. But, as in \cite{ncr},
we do not find that the relative neutrino charge radius (measured by the
photon) is gauge-invariant; the result in Ref.~\cite{bern1,bern2,bern3} is
primarily a result of the neglect in these references of sub-leading terms.

Since one motivation for studying the neutrino charge radius would be to probe
for new physics, we note that, in general, using the reaction $\nu_{\mu}+e_{R}
\rightarrow \nu_{\mu}+e_{R}$ does not simplify the analysis.  Consider models
beyond the standard model theories with strong-electroweak gauge groups
$G_{LR}=SU(3)_{c}\times SU(2)_{L}\times SU(2)_{R}\times U(1)_{B-L}$ and
$G_{422}=SU(4)_{PS}\times SU(2)_{L}\times SU(2)_{R}$, where PS stands for
Pati-Salam \cite{lrs}. Here the first step in the extraction process fails; the
$A^{\pm}_{L}$ and $A^{\pm}_{R}$ mix to form the mass eigenstates
$W_{1,2}^{\pm}$, and hence one is not able to remove the $2W$ exchange
diagrams by considering $\nu_{\mu}+e_{R} \rightarrow \nu_{\mu}+e_{R}$.  The
standard model is thus special in allowing the separation of single-particle,
i.e., $B_{\mu}$, exchange diagrams from the box diagrams.  

Neutrino masses affect neutrino electromagnetic properties, and our analysis
also applies to the gauge-dependence of both the vector neutrino charge radius
$dF_1^V(q^2)/dq^2|_{q^2=0}$ (which vanishes anyway for Majorana neutrinos) 
and the axial-vector analogue $dF_1^A(q^2)/dq^2|_{q^2=0}$. 

\bigskip
\bigskip

{\bf Acknowledgements} 
\\

We thank P. A. Grassi for useful discussions.  K.F. thanks the members of
C.N. Yang Institute for Theoretical Physics at Stony Brook for hospitality
during visits.  The research of R.S. was partially supported by the grant
NSF-PHY-00-98527. R.S. thanks the theory group of the University of Tokyo and
the organizers of the Fuji-Yoshida Summer Institute 2003 for hospitality during
visits.

\newpage

\section{Appendix}

In this appendix we display one-loop diagrams for the reaction $\nu_{\mu}+
e_{R} \to \nu_{\mu}+ e_{R}$ in the SM, in addition to the $2W$-exchange
diagram, whose contribution vanishes for $m_e \to 0$.  We show these graphs for
unitary gauge, but have analyzed the process in the full $R_\xi$ gauge.

\bigskip
\bigskip
\bigskip
\bigskip

\begin{center}
\begin{picture}(210,100)(0,0)
\ArrowLine(0,100)(200,100)
\ArrowLine(0,10)(200,10)
\Photon(60,10)(60,100){4}{8}
\Photon(150,10)(150,100){4}{8}
\Text(0,108)[]{$\nu_\mu$}
\Text(210,108)[]{$\nu_\mu$}
\Text(75,60)[]{$Z$}
\Text(165,60)[]{$Z$}
\Text(0,0)[]{$e_R$}
\Text(200,0)[]{$e_R$}
\Text(100,0)[]{(a)}
\end{picture}
\end{center}

\bigskip
\bigskip
\bigskip
\bigskip

%g14

\begin{center}
\begin{picture}(210,100)(0,0)
\ArrowLine(0,100)(200,100)
\ArrowLine(0,10)(200,10)
\Photon(60,10)(150,100){4}{8}
\Photon(150,10)(60,100){4}{8}
\Text(0,108)[]{$\nu_\mu$}
\Text(210,108)[]{$\nu_\mu$}
\Text(80,55)[]{$Z$}
\Text(130,55)[]{$Z$}
\Text(0,0)[]{$e_R$}
\Text(200,0)[]{$e_R$}
\Text(100,0)[]{(b)}
\end{picture}
\end{center}

\begin{figure}[]
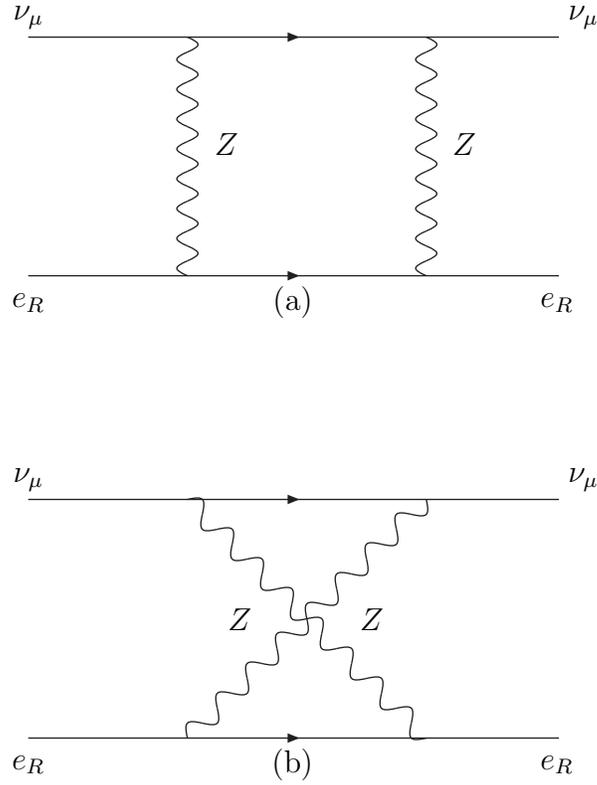

\caption{\footnotesize{$2Z$-exchange graphs.}}
\label{2Z-graphs}
\end{figure}

\newpage

\bigskip
\bigskip
\bigskip
\bigskip

\begin{center}
\begin{picture}(200,100)(0,0)
\ArrowLine(0,90)(200,90)
\ArrowLine(0,10)(200,10)
\PhotonArc(100,90)(50,0,180){4}{8.5}
\Photon(100,90)(100,10){4}{8}
\Text(0,98)[]{$\nu_\mu$}
\Text(100,98)[]{$\mu$}
\Text(200,98)[]{$\nu_\mu$}
\Text(100,150)[]{$W^+$}
\Text(113,50)[]{$\gamma$}
\Text(0,0)[]{$e_R$}
\Text(100,0)[]{(a)}
\Text(200,0)[]{$e_R$}
\end{picture}
\end{center}

\bigskip
\bigskip
\bigskip
\bigskip

%g3

\begin{center}
\begin{picture}(200,120)(0,0)
\ArrowLine(0,100)(200,100)
\ArrowLine(0,10)(200,10)
\PhotonArc(100,100)(50,180,360){4}{8.5}
\Photon(100,48)(100,10){4}{4}
\Text(0,108)[]{$\nu_\mu$}
\Text(100,108)[]{$\mu$}
\Text(200,108)[]{$\nu_\mu$}
\Text(80,70)[]{$W^+$}
\Text(113,30)[]{$\gamma$}
\Text(0,0)[]{$e_R$}
\Text(100,0)[]{(b)}
\Text(200,0)[]{$e_R$}
\end{picture}
\end{center}
\begin{center}
\begin{figure}[]
\caption{\footnotesize{Graphs contributing to $F_{\gamma}(t)$.}}
\label{F1-graphs}
\end{figure}
\end{center}

\newpage

\bigskip
\bigskip
\bigskip
\bigskip

%g4

\begin{center}
\begin{picture}(200,100)(0,0)
\ArrowLine(0,90)(200,90)
\ArrowLine(0,10)(200,10)
\PhotonArc(100,90)(50,0,180){4}{8.5}
\Photon(100,90)(100,10){4}{8}
\Text(0,98)[]{$\nu_\mu$}
\Text(100,98)[]{$\mu$}
\Text(200,98)[]{$\nu_\mu$}
\Text(100,150)[]{$W^+$}
\Text(113,53)[]{$Z$}
\Text(0,0)[]{$e_R$}
\Text(200,0)[]{$e_R$}
\Text(100,0)[]{(a)}
\end{picture}
\end{center}

\bigskip
\bigskip
\bigskip
\bigskip
\bigskip

%g5

\begin{center}
\begin{picture}(200,120)(0,0)
\ArrowLine(0,100)(200,100)
\ArrowLine(0,10)(200,10)
\PhotonArc(100,100)(50,180,360){4}{8.5}
\Photon(100,48)(100,10){4}{4}
\Text(0,108)[]{$\nu_\mu$}
\Text(100,108)[]{$\mu$}
\Text(200,108)[]{$\nu_\mu$}
\Text(80,70)[]{$W^+$}
\Text(113,30)[]{$Z$}
\Text(0,0)[]{$e_R$}
\Text(200,0)[]{$e_R$}
\Text(100,0)[]{(b)}
\end{picture}
\end{center}

\bigskip
\bigskip
\bigskip
\bigskip
\bigskip

%g6

\begin{center}
\begin{picture}(200,100)(0,0)
\ArrowLine(0,90)(200,90)
\ArrowLine(0,10)(200,10)
\PhotonArc(100,90)(50,0,180){4}{8.5}
\Photon(100,90)(100,10){4}{8}
\Text(0,98)[]{$\nu_\mu$}
\Text(100,98)[]{$\nu_\mu$}
\Text(200,98)[]{$\nu_\mu$}
\Text(100,155)[]{$Z$}
\Text(113,50)[]{$Z$}
\Text(0,0)[]{$e_R$}
\Text(200,0)[]{$e_R$}
\Text(100,0)[]{(c)}
\end{picture}
\end{center}

\begin{center}
\begin{figure}[]
\caption{\footnotesize{Graphs contributing to 
$F_Z^{\nu\nu}(t)$.}}
\label{FZnunu-graphs}
\end{figure}
\end{center}

\bigskip
\bigskip
\bigskip
\bigskip
\bigskip

%g7

\begin{center}
\begin{picture}(200,100)(0,0)
\ArrowLine(0,140)(200,140)
\ArrowLine(0,60)(200,60)
\PhotonArc(100,60)(50,180,360){4}{8.5}
\Photon(100,60)(100,140){4}{8}
\Text(0,148)[]{$\nu_\mu$}
\Text(200,148)[]{$\nu_\mu$}
\Text(115,100)[]{$Z$}
\Text(0,52)[]{$e_R$}
\Text(200,52)[]{$e_R$}
\Text(100,0)[]{$\gamma$, \ $Z$}
\Text(25,0)[]{(a)}
\end{picture}
\end{center}

\bigskip
\bigskip
\bigskip
\bigskip
\bigskip

%g8

\begin{center}
\begin{picture}(200,120)(0,0)
\ArrowLine(0,100)(200,100)
\ArrowLine(0,10)(200,10)
\PhotonArc(100,10)(50,0,180){4}{8.5}
\Photon(100,63)(100,100){4}{4}
\Text(0,108)[]{$\nu_\mu$}
\Text(200,108)[]{$\nu_\mu$}
\Text(60,60)[]{$W^-$}
\Text(113,80)[]{$Z$}
\Text(0,0)[]{$e$}
\Text(100,0)[]{$\nu_e$}
\Text(200,0)[]{$e$}
\Text(25,0)[]{(b)}
\end{picture}
\end{center}

\bigskip
\bigskip
\bigskip
\bigskip
\bigskip

%g9

\begin{center}
\begin{picture}(200,100)(0,0)
\ArrowLine(0,140)(200,140)
\ArrowLine(0,60)(200,60)
\PhotonArc(100,60)(50,180,360){4}{8.5}
\Photon(100,60)(100,140){4}{8}
\Text(0,148)[]{$\nu_\mu$}
\Text(200,148)[]{$\nu_\mu$}
\Text(115,100)[]{$Z$}
\Text(0,52)[]{$e$}
\Text(98,52)[]{$\nu_e$}
\Text(200,52)[]{$e$}
\Text(100,0)[]{$W^+$}
\Text(25,0)[]{(c)}
\end{picture}
\end{center}

\begin{figure}[]
\caption{\footnotesize{Graph (a) contributing to  
$F_Z^{ee}(t)$. For $e=e_R$, the contributions from graphs
(b) and (c) vanish in the SM.}}
\label{FZee-graphs}
\end{figure}

\newpage

\bigskip
\bigskip
\bigskip
\bigskip

%g10

\begin{center}
\begin{picture}(200,100)(0,0)
\ArrowLine(0,90)(200,90)
\ArrowLine(0,10)(200,10)
\Photon(100,50)(100,10){4}{4}
\Photon(100,90)(100,50){4}{4}
\Text(0,98)[]{$\nu_\mu$}
\Text(200,98)[]{$\nu_\mu$}
\Text(110,75)[]{$Z$}
\Text(100,50)[]{$\bullet$}
\Text(110,30)[]{$Z$}
\Text(0,0)[]{$e_R$}
\Text(200,0)[]{$e_R$}
\Text(100,0)[]{(a)}
\end{picture}
\end{center}

\bigskip
\bigskip
\bigskip
\bigskip

%g11

\begin{center}
\begin{picture}(200,100)(0,0)
\ArrowLine(0,90)(200,90)
\ArrowLine(0,10)(200,10)
\Photon(100,50)(100,10){4}{4}
\Photon(100,90)(100,50){4}{4}
\Text(0,98)[]{$\nu_\mu$}
\Text(200,98)[]{$\nu_\mu$}
\Text(110,75)[]{$Z$}
\Text(100,50)[]{$\bullet$}
\Text(110,30)[]{$\gamma$}
\Text(0,0)[]{$e_R$}
\Text(200,0)[]{$e_R$}
\Text(100,0)[]{(b)}
\end{picture}
\end{center}

\begin{figure}[]
\caption{\footnotesize{Graphs contributing to  
(a) $\Pi_{ZZ}(t)$ and (b) $\Pi_{Z\gamma}(t)$. The filled dots denote 
(a) diagonal and (b) nondiagonal vector boson propagator corrections.}}
\label{Pi-graphs}
\end{figure}


\begin{thebibliography}{99}
\bibitem{cc}
Recall that electromagnetic current conservation implies the relations 
$(m_{\nu_k}-m_{\nu_j})F_1^V(q^2)_{kj} + q^2F_3^V(q^2)_{kj} = 0$ and 
$-(m_{\nu_k}+m_{\nu_j})F_1^A(q^2)_{kj} + q^2 F_3^A(q^2)_{kj} = 0$.
\bibitem{renorm} 
In general, $F^{\nu\nu}_Z(q^{2})$ requires renormalization
and thus $[F^{\nu\nu}_Z(0)_{\nu_\mu} - 
           F^{\nu\nu}_Z(0)_{\nu_\tau} ]/M^2_Z$ 
gives the sub-leading order after renormalization.  The leading term
is absorbed into the renormalization convention of $F^{\nu\nu}_Z(q^{2})$ for
the on-shell neutrino $\nu_e$ and $q^2=0$, for example, and the difference
above becomes sub-leading.
\end{thebibliography}
\end{document}